\documentstyle[epsfig,epsf]{article}
\begin{document}

\begin{center}
{\Large {\bf Fragmentation studies of high energy ions using CR39 nuclear track detectors}}
\end{center}

\vskip .7 cm

\begin{center}
V. Togo$^a$,\footnote{Corresponding author, togo@bo.infn.it},
S. Balestra$^a$,  
S. Cecchini$^{a,b}$, 
D. Di Ferdinando$^a$,
M. Frutti$^a$, 
G. Giacomelli$^a$,  
M. Giorgini$^a$, \par 
A. Kumar$^{a,c}$,
G. Mandrioli $^a$,
S. Manzoor$^{a,d}$,     
A. Margiotta$^a$,  
E. Medinaceli$^a$,  
L. Patrizii$^a$,  
V. Popa$^{a,e}$, and 
M. Spurio$^a$
 \par~\par

{\it  $^a$Phys. Dept. of the University of Bologna and INFN, Sezione di 
Bologna, Viale C. Berti Pichat 6/2, I-40127 Bologna, Italy \\
$^b$INAF/IASF Sezione di Bologna, 40129, Bologna Italy \\
$^c$Dept. Of Physics, Sant Longowal Institute of Eng. and Tech., Longowal 
148 106 India \\ 
$^d$PRD, PINSTECH, P.O. Nilore, Islamabad, Pakistan \\
$^e$Institute of Space Sciences, Bucharest R-77125, Romania} 

\par~\par
{\small Presented at the 10$^{th}$ Inter. Symp. Radiat. Phys., Coimbra,
Portugal, 17-22 Sept. 2006.}

\vskip .7 cm
{\large \bf Abstract}\par
\end{center}

{\normalsize  We report on the measurements of the total charge changing 
fragmentation cross sections in high-energy nucleus-nucleus collisions using 
Fe, Si and Pb incident ions. Several stacks of CR39 nuclear track detectors 
with different target combinations were exposed at normal incidence to high 
energy accelerator beams to integrated densities of about 2000 ions/cm$^2$. 
The 
nuclear track detector foils were chemically etched, and ion tracks were 
measured using an automatic image analyser system. The cross section 
determination is based on the charge identification of beam ions and their 
fragments and on the reconstruction of their path through the stacks.\\

 }

\large
\section{Introduction}\label{sec:intro} Fragmentation studies of high energy 
ions are relevant for nuclear physics, cosmic ray physics,  astrophysics and 
applied physics [1]. High energy heavy ion fragmentation cross-sections are also 
useful to describe the effects of primary cosmic radiation hitting spacecraft 
walls. Important applications of the propagation of fast heavy ion beams 
through matter are given in space radiation protection and in the field 
of cancer therapy [2]. \par
In this paper we present experimental results on the fragmentation of 158 A GeV 
lead ions, 1 A GeV and 0.41 A GeV iron ions and 1 A GeV silicon ions. These 
measurements are part of a series of exposures at CERN, Brookhaven National 
Laboratory and CHIBA aimed to study the response of the CR39 nuclear detector 
and to determine the fragmentation cross sections of Pb, Fe and Si ions projectiles. 
Targets of C, CR39, CH$_2$, Al, Cu and Pb were used; they were chosen to be 
thin enough to minimise multiple interactions and thick enough to produce a 
sufficient number of fragments.

\section{Experimental procedure} We exposed several stacks made of CR39 
nuclear track detectors and different targets to different energy beams at: 
CERN-SPS, 158 A GeV Pb$^{82+}$; BNL-NSRL, 1 A GeV Fe$^{26+}$ and 
Si$^{14+}$;  CHIBA, 0.41 A GeV Fe$^{26+}$. Each stack has CR39 sheets 
upstream and downstream of the target. The exposures were performed at normal 
incidence. The charged fragments produced by projectile interactions with 
target nuclei keep most of the projectile longitudinal velocity. They can be 
detected after the target in CR39 detectors. Our CR39 sheets were manufactured 
by the Intercast Europe Co. of Parma, Italy, using a specially designed line 
of production [3]. \par
The detection principle of the CR39 [4] is based on the fact that a 
through-going heavily ionising particle produces a cylindrical 
radiation-damaged region along the ion trajectory creating a ``latent track". 
This damaged region is chemically reactive and can be etched by an appropriate 
chemical treatment. As a result, an etched cone is formed on both sides of 
each detector sheet, see Fig. 1. The cones are visible under a microscope. 
After exposure, the CR39 detectors were etched for 30 h in a 6N NaOH water 
solutions at a temperature of 70 $^\circ C$. \par

\begin{figure}[!ht]
\begin{center}
\mbox{\epsfig{figure=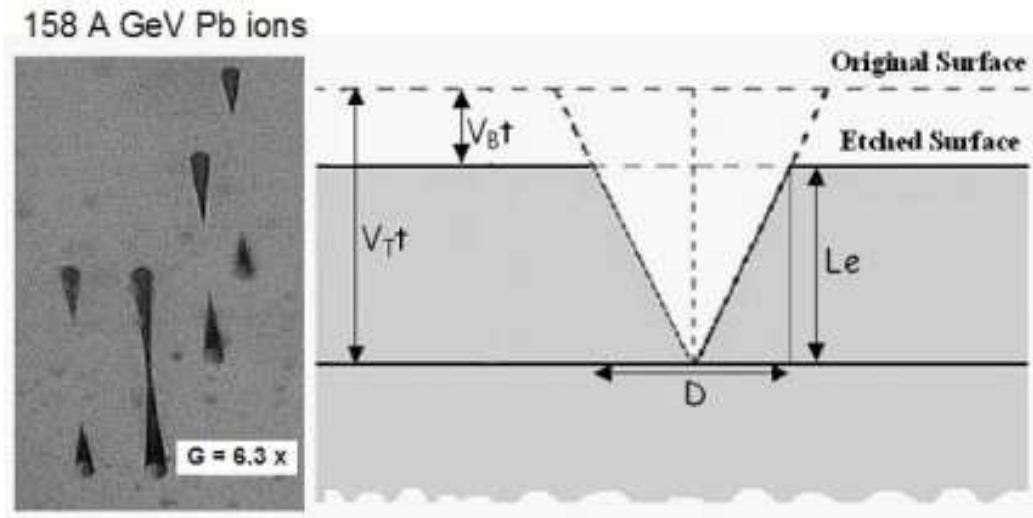,height=7 cm}}
\caption{Left: Tracks of Pb$^{82+}$ ions and their fragments in a CR39 detector; for this photomicrography the detector was inclined to show both sides. 
Right: Sketch of 
an ``etched track'' in one side of the detector for a normally incident ion. }
\label{fig:1}
\end{center}
\end{figure}

An automatic image analyser system [5] was used to scan the detector surfaces 
and measure the etch-pit cone areas. For each etch-pit cone the base area, the 
eccentricity, the central brightness and the coordinates were measured. \par
A tracking procedure was used to reconstruct the path of the beam and of the 
fragments. To better identify the projectile and fragment charges we performed 
an average of the measured etch-pit areas for each track in 3 or more sheets. Distributions of the etched cone base areas for CR39 detectors located after 
the fragmentation targets are shown in Fig. 2.  Etched cone base areas are 
given for 1 A GeV Si$^{14+}$, 1 A GeV Fe$^{26+}$ and 0.41 A GeV Fe$^{26+}$. 
Well separated 
peaks for the primary ions and for fragments are observed and a charge can be 
assigned to individual peaks; for a given z/$\beta$ value, we have the same cone 
base area for different energies (Fig. 2). \par
The reduced etch rate p=$v_T$/$v_B$, where $v_T$ and $v_B$ are the 
track and 
bulk etch velocities, respectively, was used to characterise the detector 
response [6,7]. It was determined on the basis of the surface area 
measurements of the etch-pits. The response of the detector is given by 
the relation p vs REL (Restricted Energy Loss); the REL was computed using 
the Bethe-Block formula (Particle Data Group). Fig. 3  shows the measured 
calibration curves  (p vs REL) for relativistic Pb, Fe and Si ions.

\begin{figure}[!ht]
\begin{center}
\mbox{\epsfig{figure=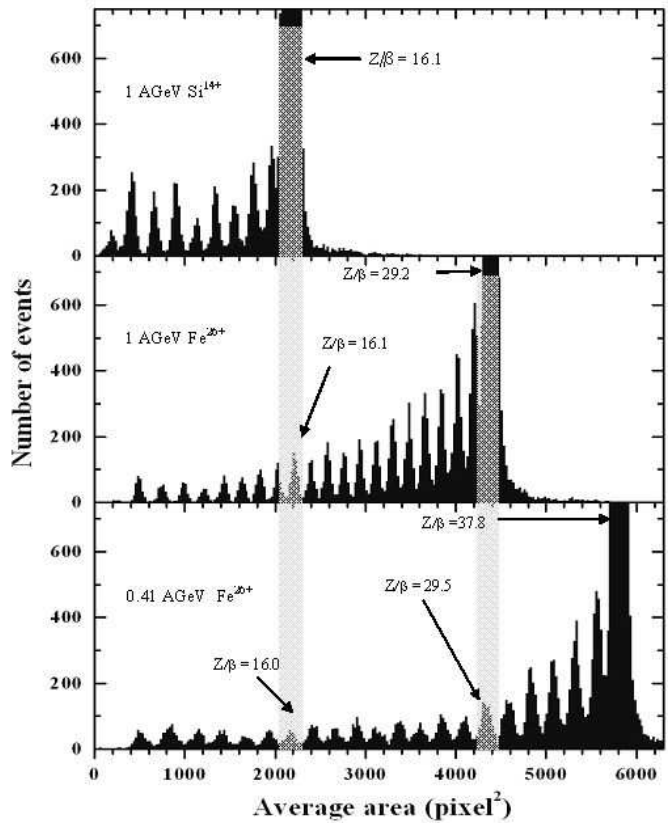,height=11 cm}}
\caption{Distributions of the etched cone areas (average areas for each track over 3 sheets) for CR39 detectors located after the fragmentation targets. Peaks for incident ions and their fragments are well separated and charges can be assigned to each peak. For a given z/$\beta$ value, we have the same cone base area for different beam energies. }
\label{fig:2}
\end{center}
\end{figure}

\vspace{-1cm}

\section{Total charge-changing cross sections}For the determination of the 
total charge-changing cross sections, $\sigma_{tot}$, the number of beam ions 
before 
the target (incident ions) and the number of beam ions after the targets were 
measured [8-11]. The target thicknesses were chosen to optimise the 
fragmentation process. \par
Our measured $\sigma_{tot}$ for the collisions of 158 A GeV Pb ions, 1 A GeV 
Fe$^{26+}$ and 
Si$^{14+}$  and 0.41 A GeV Fe$^{26+}$ on different targets are given in the 
sixth 
column of Table 1. The fragmentation charge-changing cross section for beam 
ions was evaluated using the formula

\begin{equation}
\sigma_{tot(exp)} = X_T \cdot ln(N_i~ / ~N_s) 
\end{equation}  

where $X_T = A_T/ \rho_T \cdot t_T \cdot N_A$ for each target;  $N_i$ 
is the number of primary ions, $N_S$ the number of beam ions surviving after 
the target, $\rho_T$ the target density, $A_T$ the atomic mass of the target, $t_T$  the target thickness and $N_A$ is the Avogadro number.
In this procedure, successive fragmentation processes are neglected. Hydrogen 
cross sections were obtained from the measured cross sections on carbon and on 
CH$_2$ using the formula:

\begin{equation}
\sigma_H = \frac{1} {2}(3\sigma_{CH_2}- \sigma_C)
\end{equation}

We compare our experimental cross-sections with the geometric collision cross 
section for a projectile of mass number $A_p$ on a target of mass number $A_T$:

\begin{equation}
\sigma_{tot(theo)} = \pi r_0^2(A^{1/3}_p + A^{1/3}_t - b)^2
\end{equation}

assuming $r_0$ = 1.35 fm and b = 0.83 [12]. These theoretical cross 
sections are given in the $7^{th}$ column of Table 1. 

\vspace{-0,5cm}
\begin{figure}[!ht]
\begin{center}
\mbox{\epsfig{figure=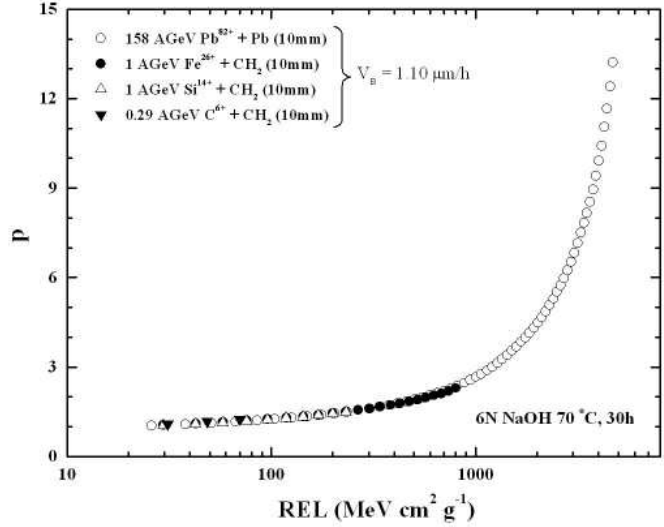,height=8.5 cm}}
\caption{p vs REL calibration curves for CR39 exposed to Lead, Iron and Silicon ions of different energies. }
\label{fig:3}
\end{center}
\end{figure}

\vspace{-1cm}

\section{Conclusions}The total charge-changing cross sections in different 
targets were measured using beams of Pb nuclei of 158 A GeV, 1 A GeV 
Fe$^{26+}$ and Si$^{14+}$, 0.41 A GeV Fe$^{26+}$ with CR39 nuclear track 
detectors placed before and after the targets, Table 1. Our results are in 
agreement with the theoretical values given by Eq. (3). \par
The calibration of the CR39 was determined by the relation p vs REL 
(Restricted Energy Loss) that shows that a unique curve gives the response 
of the detector at different energies.\par
We also exposed different stacks of CR39 to 3, 5 and 10 A GeV for both Fe and 
Si ions at the BNL AGS. These studies are in progress and should become 
available in the near future.\\

\begin{table}[!ht]
\begin{center}
\mbox{\epsfig{figure=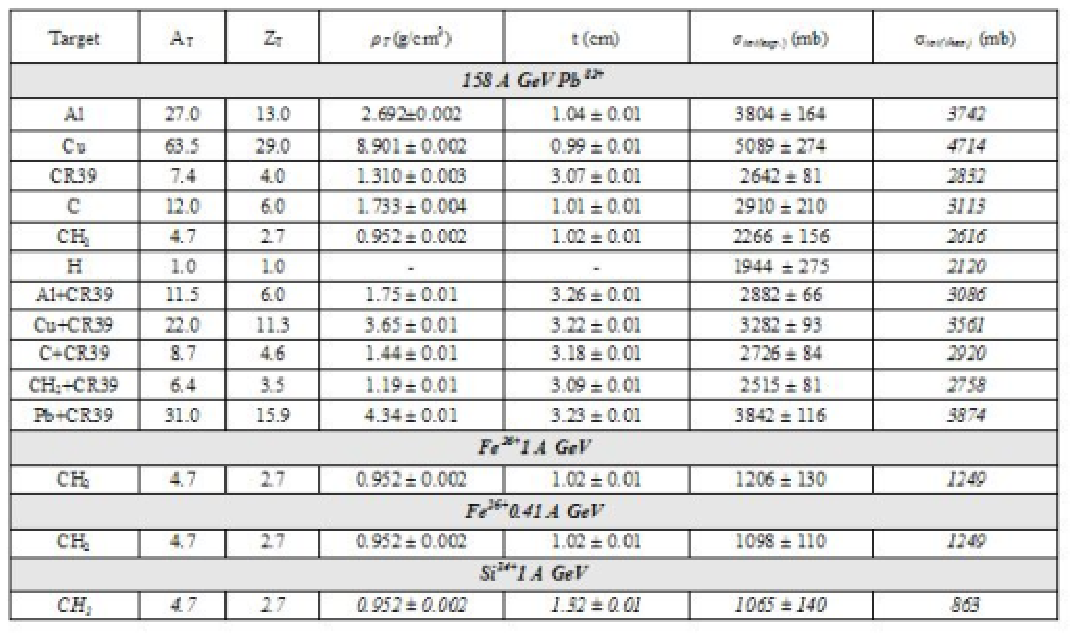,height=8 cm}}
\caption{The total charge-changing cross sections for Pb, Fe and Si ion 
projectiles on different targets. The cross sections on CH$_2$ and CR39 
are averages, as indicated in column 2. The data given in the last 3 rows are 
preliminary. The quoted uncertainties on 
$\sigma_{tot(exp.)}$ are only statistical.}
\end{center}
\end{table}

{\bf Acknowledgements.}We thank the staffs of CERN SPS, CHIBA and BNL AGS and
 NSRL for the beam exposures. We gratefully acknowledge the contributions 
of our technical staff, in  particular E.  Bottazzi, L. Degli 
Esposti, G. Grandi and C. Valieri. We thank INFN and ICTP for providing 
fellowships and grants to non-Italian citizens.

\end{document}